\documentclass[letterpaper,10pt]{article}
\usepackage{amsmath,mathptmx,graphicx,eprint}

\setreportheading{Masahiko Yoshioka}{PREPRINT}

\begin{document}

\title{Cluster synchronization in an ensemble of neurons interacting
through chemical synapses}
\author{Masahiko Yoshioka}
\date{August 22, 2003}
\author{Masahiko Yoshioka\thanks{Electric Address:
myosioka@brain.riken.go.jp} \\
{\normalsize\it Brain Science Institute, The Institute of Physical and Chemical
Research (RIKEN)}\\
{\normalsize\it Hirosawa 2-1, Wako-shi, Saitama, 351-0198, Japan}}
\date{August 22, 2003\\
{\normalsize (Revised on May 18, 2005)}}

\maketitle

\begin{abstract}
 In networks of periodically firing spiking neurons that are
 interconnected with chemical synapses, we analyze cluster state, where
 an ensemble of neurons are subdivided into a few clusters, in each of
 which neurons exhibit perfect synchronization.  To clarify stability of
 cluster state, we decompose linear stability of the solution into two
 types of stabilities: stability of mean state and stabilities of
 clusters.  Computing Floquet matrices for these stabilities, we
 clarify the total stability of cluster state for any types of neurons and any
 strength of interactions even if the size of networks is infinitely
 large.  First, we apply this stability analysis to investigating
 synchronization in the large ensemble of integrate-and-fire (IF)
 neurons.  In one-cluster state we find the
 change of stability of a cluster, which elucidates that in-phase
 synchronization of IF neurons occurs with only inhibitory synapses.  Then, we
 investigate entrainment of two clusters of IF neurons with
 different excitability. IF neurons with fast decaying synapses show the
 low entrainment capability, which is explained by a pitchfork
 bifurcation appearing in two-cluster state with change of synapse decay
 time constant. Second, we analyze one-cluster state of Hodgkin-Huxley
 (HH) neurons and discuss the difference in synchronization properties
 between IF neurons and HH neurons.
\end{abstract}

\section{Introduction}

It has been revealed that periodically firing interneurons exhibit
in-phase synchronization during  
the gamma oscillations (20-80 Hz) and the sharp wave burst (100-200
Hz)\cite{buzsaki3}.  Interneurons are found to be connected through
inhibitory chemical synapses.  Therefore, a significant effort has been
devoted to understand a role of inhibitory chemical synapses in in-phase
synchronization in a large ensemble of neurons\cite{wang}.  One major
analytical approach to investigate synchronization of neurons is the
phase reduction method, in which behavior of periodically firing neurons
are reduced to the simple phase
dynamics\cite{ermentrout,hansel,bressloff3,kuramoto}.  This phase
reduction method is, however, applicable only to the case of weak
couplings. To understand a role of strong couplings in synchronization
of neurons we have to adopt different approach.

One difficulty in investigating strongly coupled neurons is time delayed
interactions due to chemical synapses.  Taking account of these time
delayed interactions Hansel {\it et al.} have computed Floquet matrix
and analyzed synchronization in a couple of strongly coupled
neurons\cite{hansel}. The size of this Floquet matrix, however,
increases as the size of neural networks increases.  Therefore, it is
difficult to apply their approach to investigating the large size of
neural networks.

Bressloff {\it et al.} have presented another scheme to deal with
chemical synapses, which allows us to analyze stability of networks of
integrate-and-fire (IF) neurons without computing the explicit form of
Floquet matrix\cite{bressloff3}.  In some large size of neural networks
they have found the degeneracy of eigenvalues, which makes it easy to
analyze synchronization of many IF neurons.  Actually, such degenerate
eigenvalues in stability analysis are found not only in IF neurons
but also in many synchronization phenomena induced by mean field
interactions.  A most prominent example of this degeneracy is seen in
synchronization in an ensemble of chaotic oscillators such like Lorenz
equations and logistic maps\cite{fujisaka,kaneko,maistrenko,pikovsky}.
Just using the general properties of mean field
interactions we can decompose linear stability of synchronous state of
chaotic oscillators to two different components, which define the so
called tangential Lyapunov exponents and transversal Lyapunov exponents.
It must be noted that the result of this decomposition clearly indicates
the occurrence of degeneracy regarding transversal Lyapunov exponents.
Synchronization in many chaotic oscillators is thus characterized by
only a small number of exponents included in tangential and transversal
Lyapunov spectrum even if the system size is infinitely large.

In the present paper we employ these sophisticated reduction techniques
in the chaos synchronization theory to investigate synchronization in
the large number of neurons. The target of the analysis is cluster
state, where an ensemble of neurons are subdivided into a few clusters,
in each of which neurons exhibit perfect synchronization. To evaluate
the degeneracy of eigenvalues we carry out the above-mentioned
decomposition of a linear stability and define stability of
mean state (tangential Floquet multipliers) and stabilities of clusters
(transversal Floquet multipliers).  Stability of mean state elucidates
if cluster state is stable in the dynamics among clusters while
stabilities of clusters clarify whether small perturbations in each
cluster converge to vanish. We explicitly compute Floquet matrices of
these stabilities for arbitrary neuron dynamics. Therefore, we can
elucidate stability of cluster state for any types of neurons, even if
the size of networks is infinitely large and neurons are connected
through strong couplings.

To give a concrete example of the present stability analysis we first
analyze networks of IF neurons interacting through uniform chemical
synapses.  In this analysis, we find the change
of stability of a cluster, which elucidates that in-phase synchronization
of a large ensemble of IF neurons occurs with only inhibitory chemical
synapses.  In addition, we investigate two clusters of neurons with
different excitability, and discuss the relationship of their
entrainment properties to the synapse decay time constant. Second, we
analyze one-cluster state of Hodgkin-Huxley (HH) neurons and discuss the
difference in synchronization condition between IF neurons and HH
neurons.

The paper is organized as follows.  In Sec.~\ref{sec:dynamics} we
present the dynamics of neural networks that include $Q$~clusters of
spiking neurons.  In Sec.~\ref{sec:analysis}, we present the
analysis for cluster state of the neural networks.  This analysis is
applied to networks of IF neurons in Sec.~\ref{sec:if}. Then, we analyze
synchronization of HH neurons in Sec.~\ref{sec:hodgkin}. Finally, in
Sec.~\ref{sec:discussion}, we give a brief summary and discuss the
future problems that can be solved by the present approach.

\section{Networks of spiking neurons coupled with chemical synapses}\label{sec:dynamics}

We consider a spiking neuron whose state is represented by 
$n$-dimensional vector
\begin{equation}
 {{\bf x}}={\left ( {v,w_{1},w_{2},\ldots,w_{n-1}} \right )}^{{\tiny\mbox{T}}},
\end{equation}
where $v$ represent the membrane potential and
${\left\{ {w_l} \right \}}_{l=1,\ldots,n-1}$ describe gating of ion channels.
Typically, the dynamics of a spiking neuron is defined by Hodgkin-Huxley
(HH) equations, FitzHugh-Nagumo (FN) equations, and so on.  We simply
represent these neuron dynamics by
\begin{equation}
 \dot{{{\bf x}}}={\bf F}{\left ( {{{\bf x}}} \right )}.\label{eq:general}
\end{equation}
In the analysis in Sec.~\ref{sec:analysis}, we assume spiking neurons in
the form of Eq.~(\ref{eq:general}).  Nevertheless, in Sec.~\ref{sec:if},
we will investigate integrate-and-fire (IF) neurons, which cannot be
expressed by Eq.~(\ref{eq:general}) since $v$ of IF neuron changes
discontinuously.  This discontinuity of IF neuron requires a minor
corrections of the analysis in Sec.~\ref{sec:analysis}.  We will discuss
this minor correction in Sec.~\ref{sec:if}.

We assume that $N$ spiking neurons ${\left\{ {{{\bf x}}_i} \right \}}$ are interconnected
through chemical synapses.  To describe the dynamics of synaptic
electric currents, we define spike timing by the time when membrane
potential $v_i=[{{\bf x}}_i]_1$ (the first element of vector ${{\bf x}}_i$) exceeds
the threshold value $\theta=0$.  We represent $k$-th spike timing of
neurons~$i$ by $t_i(k)$, which satisfies
\begin{equation}
 v_i[t_i(k)]={\left [ {{{\bf x}}_i[t_i(k)]} \right ]}_1=\theta\label{eq:spiketiming}
\end{equation}
and
\begin{equation}
 \dot{v}_i[t_i(k)]={\left [ {\dot{{{\bf x}}}_i[t_i(k)]} \right ]}_1>0.
\end{equation}
Then, the dynamics of
networks of spiking neurons is expressed as
\begin{equation}
 \dot{{{\bf x}}}_i={\bf F}_i({{\bf x}}_i)+(I_i,0,\ldots,0)^{{\tiny\mbox{T}}}, \label{eq:dynamics}
\end{equation}
where function~${\bf F}_i({{\bf x}}_i)$ represents the dynamics of
neuron~$i$. Variable $I_i$ represents a sum of synaptic electric currents, which is defined by
\begin{equation}
I_i=\sum_{j=1}^N J_{ij}\sum_{k=-\infty}^\infty S{\left [ {t-t_j(k)} \right ]}\label{cur},
\end{equation}
where $J_{ij}$
represents synaptic coupling from neuron~$j$ to neuron~$i$, and 
function~$S(t)$ describes
time evolution of synaptic electric current.
We assume $S(t)$ taking the form
\begin{equation}
 S(t)=\left\{\begin{array}{lc}
   \displaystyle 0 & t<0,\\
   \displaystyle \frac{1}{\tau_1-\tau_2}{\left ( {e^{-t/\tau_1}-e^{-t/\tau_2}} \right )} & 0\le t .
  \end{array}
\right . \label{eq:synapse}
\end{equation}
where $0<\tau_2<\tau_1$.
Constants $\tau_1$ and $\tau_2$ are termed decay time and rise time,
respectively.

\subsection{Neural networks composed of $Q$ clusters of neurons}

In some problems, we have to consider neural networks including several
clusters of neurons, such like networks including both interneurons and
pyramidal neurons. Moreover, we will later study entrainment of two
clusters of IF neurons that have different excitability between
clusters.  In the present study we analyze neural networks that are
composed $Q$ clusters of neurons.  We assume that neurons share the same
dynamical properties within each cluster, that is, we assume
\begin{equation}
 {\bf F}_i({{\bf x}})={\bf F}_q({{\bf x}}),\quad i\in U_q,\ 1\le q \le Q,\label{eq:generalf}
\end{equation}
where $U_q$ represents the set of neurons that belong to cluster~$q$.
In addition, we assume that synaptic couplings~$J_{ij}$ depend 
only on
cluster indexes of pre and postsynaptic neurons, that is, 
we assume synaptic coupling $J_{ij}$ of the form
\begin{equation}
J_{ij}=\tilde{J}_{qq'}/N,\quad i\in U_q,\ j\in U_{q'}.\label{generalj}
\end{equation}
Substituting Eqs.~(\ref{eq:generalf}) and (\ref{generalj}) into
Eqs.~(\ref{eq:dynamics}) and (\ref{cur}) we obtain the dynamics of
$Q$~clusters of neurons:
\begin{gather}
 \dot{{{\bf x}}}_i={\bf F}_q({{\bf x}}_i)+(I_q,0,\ldots,0)^{{\tiny\mbox{T}}}, \label{eq:clusterx}\\
 I_q=\frac{1}{N}\sum_{q'=1}^Q\sum_{j\in U_{q'}}
 \tilde{J}_{qq'}\sum_k S{\left [ {t-t_j(k)} \right ]},\quad i\in U_q.\label{eq:clusteri}
\end{gather}
Note that synaptic electric current in Eq.~(\ref{eq:clusteri})
depends only on cluster
index~$q$ because of the assumption in Eq.~(\ref{generalj}).

\section{Analysis}\label{sec:analysis}

\subsection{Cluster synchronization of periodically firing neurons}

In the present analysis we focus on 
cluster state, in which spike timing of neurons are written in the form
\begin{gather}
t_i^\ast(k)=t_q^\ast(k)=t_q^\ast+kT,\nonumber\\
0\le t_q^\ast<T,\ i\in U_q,\ q=1,\ldots,Q, \label{focus}
\end{gather}
where asterisks indicates the quantity in stationary state.
In this state, neurons emit periodic spikes synchronously within each cluster.
We further assume that in cluster state not only spike timing but also
neuron states are synchronized within each
cluster (i.e., ${{\bf x}}_i^\ast={{\bf x}}_q^\ast\ (i\in U_q)$).
Substituting Eq.~(\ref{focus})
into Eqs.~(\ref{eq:clusterx}) and (\ref{eq:clusteri}),
we obtain the
dynamics of stationary state as
\begin{gather}
\dot{{{\bf x}}}_q^\ast={\bf F}_q({{\bf x}}_q^\ast)+(I_q^\ast,0,\ldots,0)^{{\tiny\mbox{T}}}, \label{sdynv}\\
I_q^\ast=\sum_{q'} \tilde{J}_{qq'}r_{q'}\tilde{S}{\left ( {t-t_{q'}^\ast} \right )},\label{scur}
\end{gather}
where $\tilde{S}(t)$ is defined by $\tilde{S}(t)=\sum_k S(t+kT)$ and $r_q=N_q/N$ represents the ratio of the number of neurons in
cluster~$q$ to the total number of neurons.

To obtain the explicit form of cluster state we have to
calculate $T$ and $t_1^\ast,t_2^\ast,\ldots,t_q^\ast$ so as to 
obtain $I_q^\ast$ and ${{\bf x}}_q^\ast$.
It is obvious that we can safely assume $t_{1}^\ast=0$, and  we can calculate remaining $Q$~unknown 
variables:~$T,t_{2}^\ast,t_{3}^\ast,\ldots,t_{Q}^\ast$ 
from {{{Eqs.}~{(\ref{sdynv})}}} and {{(\ref{scur})}} following the same scheme as our previous
study\cite{myosioka5,myosioka6}.
Note that we can compute these variables not only for IF neurons but
also for general neuron dynamics, as far as the stable
cluster state is concerned.

\subsection{Decomposition of linear stability}

To investigate linear stability of cluster state we assume the
infinitesimal deviations of state of neurons:
\begin{equation}
 {{\bf x}}_i={{\bf x}}_q^\ast+\delta{{\bf x}}_i,\quad i\in U_q\label{eq:dx}
\end{equation}
and infinitesimal deviations of spike timing:
\begin{equation}
 t_i(k)=t_{q}^\ast(k)+\delta
t_i(k),\quad i\in U_{q}.\label{eq:dt}
\end{equation}
From Eq.~(\ref{eq:spiketiming}), we obtain
\begin{gather}
\delta t_i(k)=-\delta v_i[t_{q}^\ast(k)]/c_{q}=-{\left [ {\delta{{\bf x}}_i[t_{q}^\ast(k)]} \right ]}_1/c_{q},\nonumber\\
 i\in U_{q}\label{eq:deltat}
\end{gather}
with
\begin{equation}
 c_{q}=\dot{v}_{q}^\ast[t_{q}^\ast(k)]={\left [ {\dot{{{\bf x}}}_q^\ast[t_{q}^\ast(k)]} \right ]}_1.
\end{equation}
Note that constant $c_{q}$ is independent of $k$ because of the periodicity of the solution.
To obtain the relation in Eq.~(\ref{eq:deltat}), we must assume
continuous neuron dynamics such as HH neurons and FN neurons.
Note that we have to carry out the more
careful calculation in discontinuous dynamics like IF neuron as we will discuss in Sec.~\ref{sec:if}.
Expanding the dynamics in Eqs. (\ref{eq:clusterx}) and (\ref{eq:clusteri})
to the first order we obtain  
the dynamics for the deviations:
\begin{gather}
 \delta\dot{{{\bf x}}}_i={\bf F}^\prime_q({{\bf x}}_q^\ast) \delta{{\bf x}}_i+(\delta I_q,0,\ldots,0)^{{\tiny\mbox{T}}}\label{eq:individualx},\\
 \delta I_q=-\frac{1}{N}\sum_{q'}\sum_{j\in
 U_{q'}}\tilde{J}_{qq'}\sum_k S'{\left [ {t-t_{q'}^\ast(k)} \right ]}\delta t_j(k),\label{eq:individuali}
\end{gather}
where ${\bf F}^\prime_q({{\bf x}}_q^\ast)$ denotes Jacobi matrix.

The naive evaluation of this $N\times n$-dimensional dynamics yields
an eigenvalue problem of the large size of matrix.
Therefore, for each cluster, we define mean state of neurons:
\begin{equation}
 \overline{{{\bf x}}}_q=\frac{1}{N_q}\sum_{i\in U_q}{{\bf x}}_i
\end{equation}
and mean spike timing:
\begin{equation}
 \overline{t}_q(k)=\frac{1}{N_q}\sum_{i\in U_q}t_i(k).
\end{equation}
Noting Eqs.~(\ref{eq:individualx}), (\ref{eq:individuali}), and (\ref{eq:deltat}), we can write the dynamics for
$\delta\overline{{{\bf x}}}_q$ and $\delta\overline{t}_q(k)$
in the closed form
\begin{gather}
 \delta\dot{\overline{{{\bf x}}}}_q={\bf F}^\prime_q({{\bf x}}_q^\ast)\delta\overline{{{\bf x}}}_q+(\delta
 I_q,0,\ldots,0)^{{\tiny\mbox{T}}},\label{eq:meanf}\\
\delta I_{q}=-\sum_{{q}'}\tilde{J}_{{q}{q}'}r_{{q}'}\sum_k S'{\left [ {t-t_{q'}^\ast(k)} \right ]}\delta \overline{t}_{q'}(k)\label{eq:meani}
\end{gather}
with
\begin{equation}
 \delta \overline{t}_{q}(k)=-\delta\overline{v}_{q}[t_{q}^\ast(k)]/c_{q}=-{\left [ {\delta\overline{{{\bf x}}}[t_{q}^\ast(k)]} \right ]}_1/c_{q}.\label{eq:meant}
\end{equation}
Eqs.~(\ref{eq:meanf})-(\ref{eq:meant}) define the decomposed stability of the original $N$-body
stability.
We term this decomposed stability stability of mean state.
It must be noted that 
stability of mean state in
Eqs.~(\ref{eq:meanf})-(\ref{eq:meant}) is effectively a problem in a
network of $Q$ neurons 
with couplings $J_{{q}{q}'}r_{{q}'}$ since, to the first order,
Eqs.~(\ref{eq:meanf})-(\ref{eq:meant}) are equivalent to 
\begin{gather}
 \frac{d}{dt}{\left ( {{{\bf x}}_q^\ast+\delta\overline{{{\bf x}}}_q} \right )}={\bf F}_q{\left ( {{{\bf x}}_q^\ast+\delta\overline{{{\bf x}}}_q} \right )}+(I_q,0,\ldots,0)^{{\tiny\mbox{T}}},\label{eq:unperturbedmeanx}\\
 I_q=\sum_{q'}\tilde{J}_{qq'}r_{q'}\sum_kS{\left [ {t-t_q^\ast(k)-\delta\overline{t}_{q'}(k)} \right ]}.\label{eq:unperturbedmeani}
\end{gather}

Stability of mean state is a necessary condition for the
full stability, but not a sufficient condition.
To investigate synchronization of neurons in each cluster we
introduce deviations around the averaged state:
\begin{equation}
 {{\bf x}}_i=\overline{{{\bf x}}}_q+\delta\tilde{{{\bf x}}}_i,\quad i\in U_q.
\end{equation}
Subtracting Eq.~(\ref{eq:meanf}) from Eq.~(\ref{eq:individualx}) we obtain
the dynamics of $\delta\tilde{{{\bf x}}}_i$ as
\begin{gather}
 \delta\dot{\tilde{{{\bf x}}}}_i={\bf F}^\prime_q({{\bf x}}_q^\ast)\delta{\tilde{{{\bf x}}}}_i,\quad i\in U_q.\label{eq:tildex}
\end{gather}
Eq~(\ref{eq:tildex}) defines another decomposed
stability.
We term this decomposed stability stability of a cluster.
Stability of cluster~${q}$ is satisfied when
$N_q$ deviations $\delta\tilde{{{\bf x}}}_i\ (i\in
U_{q})$ converge into ${\bf 0}$.
These $N_q$ dynamics are, however, identical.
Therefore, it suffices to evaluate one set of
deviations to
determine the stability of one cluster.
Note that the determination
of the stability of a
cluster is effectively a problem of a single neuron dynamics 
under the unperturbed synaptic electric current~$I_q^\ast$ since, to the
first order, Eq.~(\ref{eq:tildex}) is equivalent to
\begin{equation}
 \frac{d}{dt}{\left ( {{{\bf x}}^\ast_q+\delta\tilde{{{\bf x}}}_i} \right )}={\bf F}_q({{\bf x}}_q^\ast+\delta\tilde{{{\bf x}}}_i)+(I_q^\ast,0,\ldots,0)^{{\tiny\mbox{T}}},\quad i\in U_q.\label{eq:efftildex}
\end{equation}

\subsection{Floquet matrices for stabilities of clusters}\label{sec:floquet-cluster}

We can determine 
stabilities of clusters following the ordinary procedure of Floquet theory.
Since the solution~${{\bf x}}^\ast_q$ is periodic,
${\bf F}^\prime_q({{\bf x}}_q^\ast)$ is also periodic.
Therefore, a solution of Eq.~(\ref{eq:tildex}) is written in the form
\begin{equation}
 \delta\tilde{{{\bf x}}}_i[t_q^\ast(k+1)]={\bf M}^{{{\tiny \perp}}}_q\delta\tilde{{{\bf x}}}_i[t_q^\ast(k)],\quad i\in U_q.
\end{equation}
Calculating Eq.~(\ref{eq:efftildex}) with small initial
perturbations
we can obtain every elements in matrix~${\bf M}_q^{{\tiny \perp}}$.
$n\times n$~matrix ${\bf M}_q^{{\tiny \perp}}$ has $n$~eigenvalues
${\left\{ {\lambda_{ql}^{{\tiny \perp}}} \right \}}_{l=1,\ldots,n}$.
When cluster~$q$ is stable, 
$\delta\tilde{{{\bf x}}}_i\ (i\in U_q)$ must converge to zero after a long time.
Therefore, the stability of cluster~$q$ is fulfilled when the largest
absolute eigenvalue~$|\lambda_{q1}^{{\tiny \perp}}|$ satisfies the condition 
\begin{equation}
 |\lambda_{q1}^{{\tiny \perp}}| < 1.\label{eq:stabilityofclusters}
\end{equation}

\subsection{Floquet matrix for stability of mean state}

Determination of the stability of mean state is not an easy problem
since the calculation of $\delta I_{q}$ in {{{Eq.}~{(\ref{eq:meani})}}} requires long past
deviations of spike timing.
To solve this problem, following Hansel~{\it et al.}\cite{hansel}, we introduce
the variables:
\begin{eqnarray}
z_{{q}1}&=&\sum_{{q}'}\tilde{J}_{{q}{q}'}r_{{q}'}\sum_{t_{{q}'}^\ast(k')<t}S{\left [ {t-t_{{q}'}^\ast(k')-\delta
\overline{t}_{{q}'}(k')} \right ]}\label{eq:z1}\\
 z_{{q}2}&=&\sum_{{q}'}\tilde{J}_{{q}{q}'}r_{{q}'}
\sum_{t_{{q}'}^\ast(k')<t}e^{-{\left [ {t-t_{{q}'}^\ast(k')-\delta\overline{t}_{{q}'}(k')} \right ]}/\tau_1}.\label{eq:z2}
\end{eqnarray}
By means of these variables we can exactly rewrite $I_{q}$ in Eq.~(\ref{eq:unperturbedmeani}) in the
truncated form
\begin{eqnarray}
I_{q}&=&\sum_{{q}'}\tilde{J}_{{q}{q}'}r_{{q}'}\sum_{t_{q}^\ast(k) \le
t_{{q}'}^\ast(k') }
S{\left [ {t-t_{{q}'}^\ast(k')-\delta\overline{t}_{{q}'}(k')} \right ]}\nonumber\\
&&+e^{-{\left [ {t-t_{q}^\ast(k)} \right ]}/\tau_2}z_{{q}1}[t_{q}^\ast(k)]\nonumber\\
&&+S{\left [ {t-t_{q}^\ast(k)} \right ]} z_{{q}2}[t_{q}^\ast(k)],\qquad t_{q}^\ast(k)< t.\label{eq:truncated}
\end{eqnarray}
Therefore, $\delta I_{q}$ is written as
\begin{eqnarray}
\delta I_{q}&=&-\sum_{{q}'}\tilde{J}_{{q}{q}'}r_{{q}'}\sum_{t_{q}^\ast(k) \le
t_{{q}'}^\ast(k') }
S'{\left [ {t-t_{{q}'}^\ast(k')} \right ]}\delta\overline{t}_{{q}'}(k')\nonumber\\
&&+e^{-{\left [ {t-t_{q}^\ast(k)} \right ]}/\tau_2}\delta
 z_{{q}1}[t_{q}^\ast(k)]\nonumber\\
&&+S{\left [ {t-t_{q}^\ast(k)} \right ]}\delta z_{{q}2}[t_{q}^\ast(k)],\qquad t_{q}^\ast(k)< t.\label{eq:truncateddelta}
\end{eqnarray}
This means that once we know $\delta z_{{q}1}[t_{q}^\ast(k)]$ and
$\delta z_{{q}2}[t_{q}^\ast(k)]$, we can neglect past deviations of spike
timing $\delta\overline{t}_{q'}(k')$ that arose before $t=t_{q}^\ast(k)$.

To take the advantage of $z_{{q} 1}$ and $z_{{q}2}$, we define the vector
\begin{equation}
 {\bf y}_{q}={\left ( {[\overline{{{\bf x}}}_q]_1,\ldots,[\overline{{{\bf x}}}_q]_n,z_{q1},z_{q2}} \right )}^{{\tiny\mbox{T}}}.
\end{equation}
We safely assume $t_q^\ast\le t_{q+1}^\ast\ (q=1,\ldots,Q-1)$.  
Then, since Eq.~(\ref{eq:meani}) is equivalent to
Eq.~(\ref{eq:truncateddelta}), from Eqs.~(\ref{eq:meanf}),
(\ref{eq:truncateddelta}), and (\ref{eq:meant}) we can show that
deviation~$\delta{\bf x}_q[t_q^\ast(k+1)]$ is determined from only
$\delta{\bf y}_q[t_q^\ast(k)]$ and $\delta\overline{t}_{q'}(k_{qq'})$
with
\begin{equation}
 k_{qq'}=\left \{\begin{array}{cc}
     k& q\le q',\\
     k+1  & q'<q.
	    \end{array}
\right . \label{eq:defkprime}
\end{equation}
Moreover, deviations~$\delta
z_{q1}[t_q^\ast(k+1)]$ and $\delta z_{q2}[t_q^\ast(k+1)]$ are given as
 \begin{eqnarray}
&&\delta z_{q1}[t_q^\ast(k+1)]\nonumber\\
 &=&-\sum_{q'}\tilde{J}_{{q}{q}'}r_{{q}'}
S'{\left [ {t_q^\ast(k+1)-t_{{q}'}^\ast(k_{qq'})} \right ]}\delta\overline{t}_{{q}'}(k_{qq'})\nonumber\\
&&+e^{-T/\tau_2}\delta\label{eq:deltaz1}
 z_{{q}1}[t_{q}^\ast(k)]+S{\left ( {T} \right )}\delta z_{{q}2}[t_{q}^\ast(k)],\\
&&\delta z_{q2}[t_q^\ast(k+1)]\nonumber\\
 &=&\frac{1}{\tau_1}\sum_{q'}\tilde{J}_{{q}{q}'}r_{{q}'}
e^{-{\left [ {t_q^\ast(k+1)-t_{{q}'}^\ast(k_{qq'})} \right ]}/\tau_1}\delta\overline{t}_{{q}'}(k_{qq'})\nonumber\\
  &&+e^{-T/\tau_1} \delta z_{q2}[t_q^\ast(k)].\label{eq:deltaz2}
 \end{eqnarray}
Therefore, we can also determine $\delta z_{q1}[t_q^\ast(k+1)]$ and
$\delta z_{q2}[t_q^\ast(k+1)]$ from the above
mentioned variables: $\delta{\bf
y}_q[t_q^\ast(k)]$ and  $\delta\overline{t}_{q'}(k_{qq'})$.
We can summarize these relationships in the form
\begin{eqnarray}
\delta{\bf y}_q[t_q^\ast(k+1)]
 &=&\sum_{{q}'=1}^{Q}{\bf A}_{{q}{q}'}\delta{\bf
y}_{{q}'}[t_{q'}^\ast(k_{qq'})]
+{\bf B}_{q} \delta{\bf y}_{q}[t_{q'}^\ast(k)],\nonumber\\
 &&\label{eq:relationship}
\end{eqnarray}
where
\begin{equation}
 {\bf A}_{qq'}=
		\left ( \begin{array}{cccc}
  {\displaystyle \frac{\partial \delta{\bf y}_q[t_q^\ast(k+1)]}{\partial
   \delta\overline{t}_{q'}(k_{qq'})}{\left ( {-\frac{1}{c_{q'}}} \right )}}&{\bf 0}&\ldots&{\bf 0}
			\end{array}\right)
\end{equation}
and
\begin{equation}
{\bf B}_q=
		\left ( \begin{array}{ccc}
  {\displaystyle
   \frac{\partial \delta{\bf y}_q[t_q^\ast(k+1)]}{\partial [\delta{\bf
   y}_q[t_q^\ast(k)]]_1}}&
     \ldots &
   {\displaystyle \frac{\partial \delta{\bf y}_q[t_q^\ast(k+1)]}{\partial [\delta{\bf
   y}_q[t_q^\ast(k)]]_{n+2}}}
			\end{array}\right).
\end{equation}
In this equation, ${\bf A}_{qq'}\delta{\bf
y}_{q'}[t_{q'}^\ast(k_{qq'})]$ represents the contribution from $\delta\overline{t}_{q'}(k_{qq'})$.

In Sec.~\ref{sec:floquet-cluster}, we obtain ${\bf M}_q^{{\tiny \perp}}$ by
calculating Eq.~(\ref{eq:efftildex}) with small perturbations.  In the
similar manner, we can compute ${\bf A}_q$ and ${\bf B}_{qq'}$
explicitly for arbitrary neuron dynamics.  We obtain $\partial
\delta{\bf x}_q[t_q^\ast(k+1)]/\partial
\delta\overline{t}_{q'}(k_{qq'})$ and $\partial \delta{\bf
x}_q[t_q^\ast(k+1)]/\partial [\delta{\bf y}_q[t_q^\ast(k)]]_l$ by
calculating Eqs.~(\ref{eq:unperturbedmeanx}) and (\ref{eq:truncated})
with small perturbations.  Partial derivatives of $z_{q1}$ and $z_{q2}$
have been given in Eqs.~(\ref{eq:deltaz1}) and (\ref{eq:deltaz2}).
Therefore, we can compute every elements in matrices ${\bf A}_q$ and
${\bf B}_{qq'}$.  For the further details of calculation of ${\bf
A}_{q}$ and ${\bf B}_{qq'}$ see Ref.~\cite{myosioka6}(, though the
definitions of ${\bf A}_{q}$, ${\bf B}_{qq'}$, and so on in
Ref.~\cite{myosioka6} are slightly
different from the present ones.)

We introduce vector
\begin{equation}
 {\bf Y}(k)={\left ( {{\bf y}_1[t_1^\ast(k)]^{{\tiny\mbox{T}}} \ldots {\bf y}_Q[t_Q^\ast(k)]^{{\tiny\mbox{T}}}} \right )}^{{\tiny\mbox{T}}}.
\end{equation}
Then, we can rewrite the relationship in Eq.~(\ref{eq:relationship}) in the form
\begin{equation}
 \delta{\bf Y}(k+1)={\bf M}^{{\tiny \parallel}} \delta{\bf Y}(k)
\end{equation}
with
\begin{equation}
{\bf M}^{{\tiny \parallel}}={\bf M}^{{\tiny \parallel}}_{Q}{\bf M}^{{\tiny \parallel}}_{{Q}-1}\ldots {\bf M}^{{\tiny \parallel}}_1, \label{eq:mta}
\end{equation}
where
\begin{equation}
 {\bf M}^{{\tiny \parallel}}_q=\left(
\begin{array}{ccccccc}
 {\bf E}& &{\bf 0}& & & & \\
  &\ddots &      & & &{\bf 0}& \\
 {\bf 0}& &{\bf E}& & & & \\
 {\bf A}_{q1} &\ldots&{\bf A}_{qq-1} & {\bf A}_{qq}+{\bf B}_q &{\bf A}_{qq+1}
  &\ldots&{\bf A}_{qQ} \\
  & & & &{\bf E}& & {\bf 0}\\
  & {\bf 0}& & & &\ddots& \\
  & & & &{\bf 0} & &{\bf E}
\end{array}		    
		   \right).\label{eq:mtaq}
\end{equation}
Matrix~${\bf M}_{q}^{{\tiny \parallel}}$ updates $\delta {\bf y}_{q}(k)$ to $\delta
{\bf y}_{q}(k+1)$, and hence matrix~${\bf M}^{{\tiny \parallel}}$ updates all the deviations.
$Q(n+2)\times Q(n+2)$ matrix~${\bf M}^{{\tiny \parallel}}$ has $Q(n+2)$ eigenvalues
~${\left\{ {\lambda_{l}^{{\tiny \parallel}}} \right \}}_{l=1,\ldots,Q(n+2)}$, in which 
a trivial
eigenvalue one is always included as in the case of ordinary Floquet matrix.
The stability of mean state is satisfied when all other eigenvalues are
less than one in absolute value, that is, the largest absolute 
eigenvalue $|\lambda_1^{{\tiny \parallel}}|$ and the second largest absolute
eigenvalue $|\lambda_2^{{\tiny \parallel}}|$ satisfy
\begin{equation}
 \left | \lambda_2^{{\tiny \parallel}} \right | <1= \lambda_1^{{\tiny \parallel}} .
\end{equation}

\section{Cluster synchronization in networks of integrate-and-fire (IF)
 neurons}\label{sec:if}

\begin{figure}[h]
 {}{}{}
\begin{center}
\includegraphics[scale=1.5]{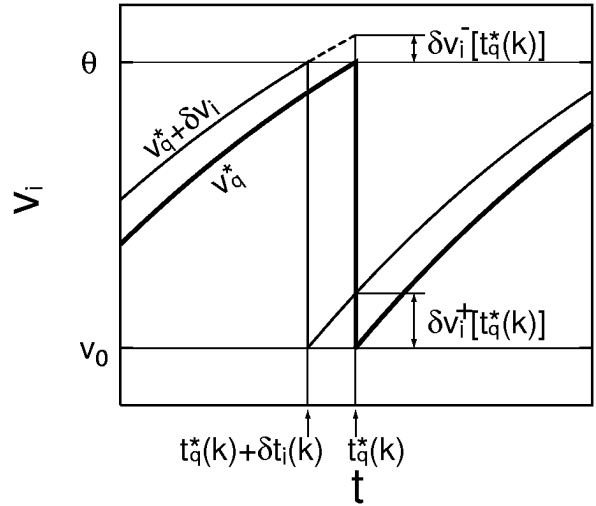}
\end{center}
 \caption{A schematic figure explaining the definition of
$\delta v_i^-[t_q^\ast(k)]$ and
$\delta v_i^+[t_q^\ast(k)]$.
Membrane potential of a IF neuron~$v_i$ changes
 discontinuously at spike timing $t_q^\ast(k)+\delta t_i(k)$.
When $t_q^\ast(k)+\delta t_i(k)<t_q^\ast(k)$, we define $\delta v_i^-[t_q^\ast(k)]$ by
 extending the solution as shown in the figure, and we define $\delta
 v_i^+[t_q^\ast(k)]=\delta v_i[t_q^\ast(k)]$.
When $t_q^\ast(k)<t_q^\ast(k)+\delta t_i(k)$, we define these variables in
 the opposite way.}\label{fig:schematic}
\end{figure}

Let us apply the above analysis to networks of IF neurons
that are defined as
\begin{equation}
\dot{v_i}=-v_i+v_r+I_{{{\mbox{\footnotesize {ext}}}},q}+I_i,\qquad i\in U_q.\label{ifv}
\end{equation}
When $v_i$ exceeds the threshold value $\theta=0$, $v_i$ is reset to
$v_0=-1$. The resting potential $v_r$ is set to $v_r=1$, which leads
intrinsic firing of neurons.
We assume that these IF neurons are interconnected with
uniform couplings:
\begin{equation}
J_{ij}=\frac{g}{N}.\label{uniform}
\end{equation}
As we have mentioned, the discontinuity 
of IF neurons require a minor correction of the 
stability analysis in Sec.~\ref{sec:analysis}.
Since derivative $\dot{v}_i$ changes discontinuously at spike timing,
we define 
$c_{{q}}^-=\dot{v}_{q}^\ast[t_{q}^\ast(k)-0]$
and $c_{{q}}^+=\dot{v}_{q}^\ast[t_{q}^\ast(k)+0]$.
To take account of discontinuity of $v_i$,  
we extend the perturbed solution $v_q^\ast+\delta v_i$ before/after spike
timing~$t_i(k)=t_q^\ast(k)+\delta t_i(k)$  as
illustrated in Fig.~\ref{fig:schematic}, and then define 
$\delta v_i^-[t_q^\ast(k)]$ and $\delta v_i^+[t_q^\ast(k)]$.
These deviations satisfy the condition,
\begin{equation}
 \delta t_i(k)=-\delta v_i^-[t_q^\ast(k)]/c_q^-=-\delta v_i^+[t_q^\ast(k)]/c_q^+.\label{eq:cplusminus}
\end{equation}
We define two types of
mean state variables:
$\delta\overline{v}_q^-=(1/N)\sum_{i\in U_q}\delta v_i^-$ and
$\delta\overline{v}_q^+=(1/N)\sum_{i\in U_q}\delta v_i^+$, and two types
of deviations around the mean state:
$\delta v_i^-=\delta\overline{v}^-_i+\delta\tilde{v}_i^-$ and $\delta
v_i^+=\delta\overline{v}^+_i+\delta\tilde{v}_i^+$.
Neuron~$i$ behaves continuously in the time interval
$t_i(k)<t<t_i(k+1)$, during which
we can carry out the decomposition of
linear stability discussed in Sec.~\ref{sec:analysis}.
Therefore, noting Eqs.~(\ref{eq:tildex}), (\ref{ifv}),
and (\ref{eq:cplusminus}), 
we obtain 
\begin{equation}
 \delta \tilde{v}_i^-[t_q^\ast(k+1)]=e^{-T}\delta \tilde{v}_i^+[t_q^\ast(k)]=\frac{c_q^+}{c_q^-}e^{-T}\delta \tilde{v}_i^-[t_q^\ast(k)].
\end{equation}
Hence, matrix~${\bf M}_q^{{\tiny \perp}}$ takes the form
\begin{equation}
 {\bf M}_q^{{\tiny \perp}}={\left ( {\frac{c_q^+}{c_q^-}e^{-T}} \right )}.
\end{equation}
From Eq.~(\ref{eq:stabilityofclusters}), we obtain the condition for
stability of cluster~$q$:
\begin{equation}
\left |\lambda_{q1}^{{\tiny \perp}}\right|=\left |\frac{c_{{q}}^+}{c_{{q}}^-}e^{-T}\right |<1.
\end{equation}

Following the similar scheme, we can derive matrices ${\bf A}_{{q}{q}'}$
and ${\bf B}_{{q}}$.  Substituting ${\bf A}_{{q}{q}'}$ and ${\bf
B}_{{q}}$ into Eqs.~(\ref{eq:mta}) and (\ref{eq:mtaq}) yields ${\bf
M}^{{\tiny \parallel}}$, by which we can determine the stability of mean state.

\subsection{One-cluster state of IF neurons ($Q=1$)}

\begin{figure}
{}{}{}
{}{}{}
{}
\begin{center}
\begin{tabular}{ccc}
{(a)} & & \ \ \ \\
 & \includegraphics[scale=1.5]{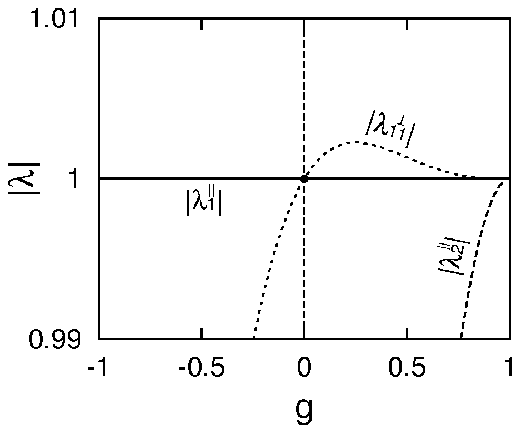} \\
{(b)} \\
 & \includegraphics[scale=1.5]{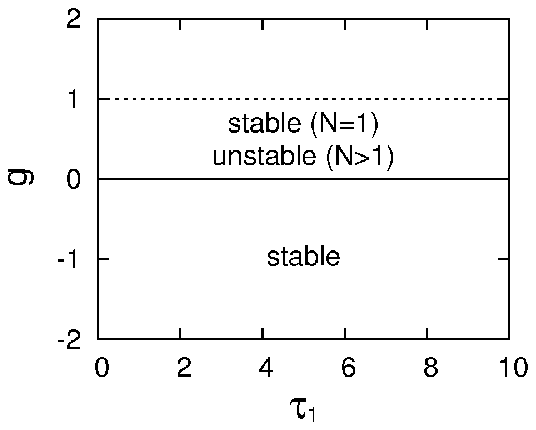} 
\end{tabular}
\end{center}
 
\caption{{(a)}\ Absolute values of $\lambda^{{\tiny \parallel}}_1$,
 $\lambda^{{\tiny \parallel}}_2$, and $\lambda^{{\tiny \perp}}_{11}$ for one-cluster state
 ($Q=1$) of networks of IF neurons are plotted as a function of $g$ for
 $I_{{\mbox{\footnotesize {ext,1}}}}=0$,$\tau_1=3.5$, and $\tau_2=0.1\tau_1$.
 $\lambda^{{\tiny \parallel}}_1$ always takes one while $|\lambda^{{\tiny \parallel}}_2|$ is
 always less than one.  $|\lambda^{{\tiny \perp}}_{11}|$ is less than one only
 when synapses are inhibitory ($g<0$).  These eigenvalues behave in the same manner even with the other decay time $\tau_1>0$.
{(b)}\ $\tau_1$-$g$~phase diagram, where we fix $\tau_2=0.1\tau_1$.
A self-coupled single neuron ($N=1$) has the stable periodic solution
below $g=1$. However, synchronization of multiple neurons ($N>1$)
occurs with only inhibitory couplings $g<0$ since the stability
of a cluster is fulfilled with only inhibitory couplings $g<0$.
Beyond $g=1$, an excessive amount of positive synaptic electric
current leads bursting of neurons.}\label{c1}
\end{figure}

We begin with investigating one-cluster state $Q=1$.  In this state, all
neurons take part in shaping one-cluster in-phase synchronization.
One-cluster solution of Eqs.~(\ref{sdynv}) and (\ref{scur}) is found
with only $g<1$ since too strong synaptic couplings with $g\ge 1$ leads
bursting of neurons.  To elucidate the stability of the solution with
$g<1$, assuming $\tau_1=3.5$, $\tau_2=0.1\tau_1$, and
$I_{{{\mbox{\footnotesize {ext}}}},1}=0$, we calculate $|\lambda_{1}^{{\tiny \parallel}}|$,
$|\lambda_{2}^{{\tiny \parallel}}|$, and $|\lambda_{11}^{{\tiny \perp}}|$ as a function of
parameter $g$ as shown in Fig.~\ref{c1}{(a)}.  Since the second largest
absolute eigenvalue of ${\bf M}^{{\tiny \parallel}}$ (i.e., $|\lambda_{2}^{{\tiny \parallel}}|$)
is always less than one, the stability of mean state is always
satisfied.  However, the largest absolute eigenvalue of ${\bf
M}_1^{{\tiny \perp}}$ (i.e., $|\lambda_{11}^{{\tiny \perp}}|$) becomes grater than one
with excitatory coupling $g>0$.  Therefore, the stability of a cluster
is satisfied with only inhibitory coupling $g<0$.  These results imply
that while a self-coupled single neuron ($N=1$) can exhibit stable
periodic firing with both inhibitory and excitatory couplings, in-phase
synchronization of multiple neurons ($N>1$) can take place with only
inhibitory couplings $g<0$.  Since the networks show the same
synchronization properties in all the decay time $\tau_1>0$,
$\tau_1-g$~phase diagram takes the simple form as described in
Fig.~\ref{c1}{(b)}.  It turns out that in-phase synchronization of a
large number of IF neurons occurs with only inhibitory synapses in all
the value $\tau_1>0$.

\begin{figure}
{}{}{}
{}{}{}
{}
\begin{center}
\begin{tabular}{ccc}
{(a)} & & \ \ \ \\
 & \includegraphics[scale=1.5]{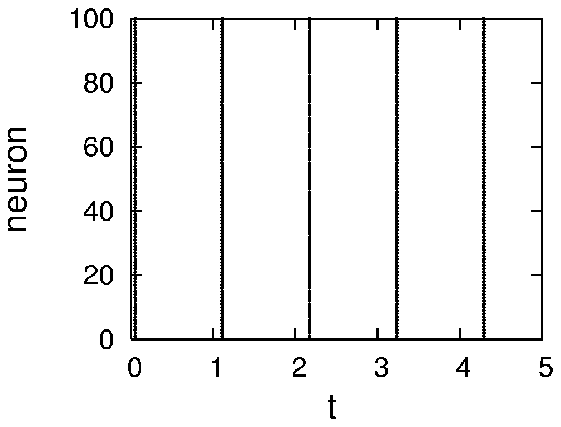} \\
{(b)} \\
 & \includegraphics[scale=1.5]{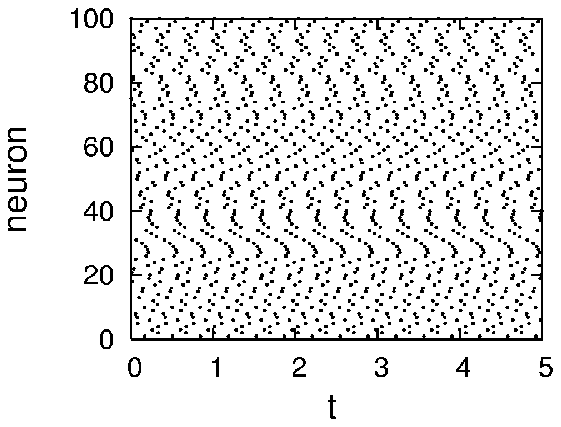} 
\end{tabular}
\end{center}
\caption{The result of numerical simulations
with $N=100$, $\tau_1=3.5$, and $\tau_2=0.1\tau_1$.
Dots represent spike timing of neurons in a stationary state, which is
 realized after a long run of simulation.
{(a)}\ With inhibitory synapses $g=-0.5$, 
the perfect in-phase synchronization occurs.
{(b)}\ With excitatory synapses $g=0.5$, we observe asynchronous state, in
 which neurons fire periodically with uniformly distributed phase shifts.}\label{fig:simulation}
\end{figure}

Figure~\ref{fig:simulation} shows the result of the numerical
simulations.  While the networks with inhibitory couplings ($g=-0.5$)
exhibits the perfect in-phase synchronization, the network with
excitatory couplings ($g=0.5$) settles into the asynchronous state, in
which neurons fire periodically with uniformly distributed phase shifts.
Our stability analysis explains these numerically results well.

\subsection{Two-cluster state of IF neurons ($Q=2$ and $I_{{{\mbox{\footnotesize {ext}}}},1}= I_{{{\mbox{\footnotesize {ext}}}},2}=0$)}

We then investigate two-cluster state $Q=2$ for inhibitory coupling
$g<0$ assuming $r_1=r_2=0.5$ and $I_{{{\mbox{\footnotesize {ext}}}},1}=I_{{{\mbox{\footnotesize {ext}}}},2}=0$. It
has been shown that a couple of IF neurons exhibit a pitchfork
bifurcation with change of synapse decay time constant\cite{hansel}.  We
now show that this pitchfork bifurcation occurs even in systems of two
clusters of neurons.  Figure~\ref{bifurcation} shows $\tau_1-\varphi_2$
bifurcation diagram, where $\varphi_2$ denotes $t_2/T$.  There are three
types of solutions: in-phase ($\varphi_2=0,1$), anti-phase
($\varphi_2=0.5$), and out-of-phase solutions.  Evaluating eigenvalues
of ${\bf M}^{{\tiny \parallel}}$, ${\bf M}_1^{{\tiny \perp}}$, and ${\bf M}_2^{{\tiny \perp}}$, we
find that the solutions denoted by thick lines satisfy the stability of
mean state and the stabilities of clusters.

\subsection{Entrainment of two clusters of IF neurons with different excitability ($Q=2$ and
  $I_{{{\mbox{\footnotesize {ext}}}},1}=0\ne I_{{{\mbox{\footnotesize {ext}}}},2}$)} 

\begin{figure}
{}{}{}
\begin{center}
\includegraphics[scale=1.5]{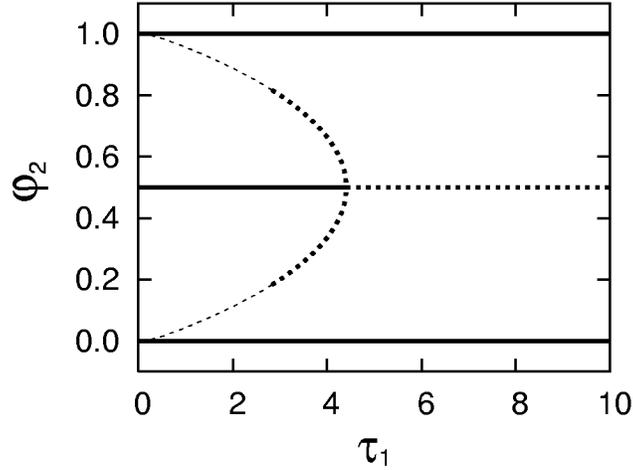}
\end{center}
\caption{$\tau_1-\varphi_2$ bifurcation diagram for two-cluster state, where variable $\varphi_2$ denotes
$t_2/T$ ($Q=2$, $1\leq N_1=N_2$, $g=-3$, $\tau_2=0.1\tau_1$, and
$I_{{{\mbox{\footnotesize {ext}}}},1}=I_{{{\mbox{\footnotesize {ext}}}},2}=0$.)
 The solutions represented by thick
lines satisfy the stability of mean state and stabilities of
clusters, while solutions represented by the dotted lines lack one or
both of stabilities.
The out-of-phase solutions plotted by the thin dotted
line ($\tau_1< 2.8$) is invalid since in these solutions
 membrane potential $v_i$ crosses the
threshold $\theta$ multiple times.}\label{bifurcation}
\end{figure}

\begin{figure}
{}{}{}
\begin{center}
\includegraphics[scale=1.5]{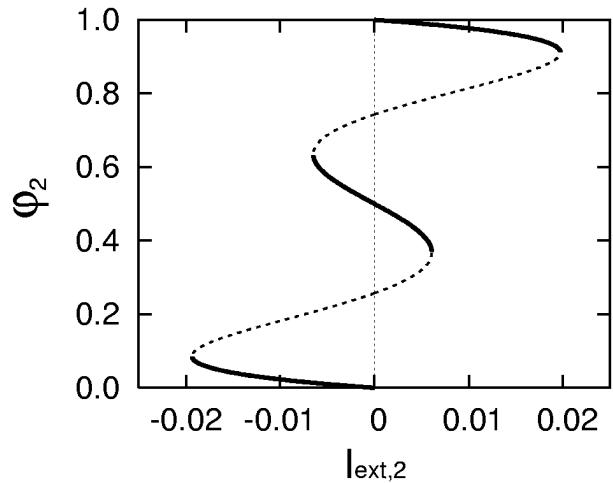}
\end{center}
\caption{Entrainment of two clusters of neurons with different
excitability.  The solution~$\varphi_2$ is plotted against
$I_{{{\mbox{\footnotesize {ext}}}},2}$ for the fixed value of $I_{{{\mbox{\footnotesize {ext}}}},1}=0$
($Q=2,1\leq N_1=N_2,g=-3,\tau_1=3.5,$ and $\tau_2=0.1\tau_1$.)}\label{8}
\end{figure}

\begin{figure}
{}{}{}
\begin{center}
\includegraphics[scale=1.5]{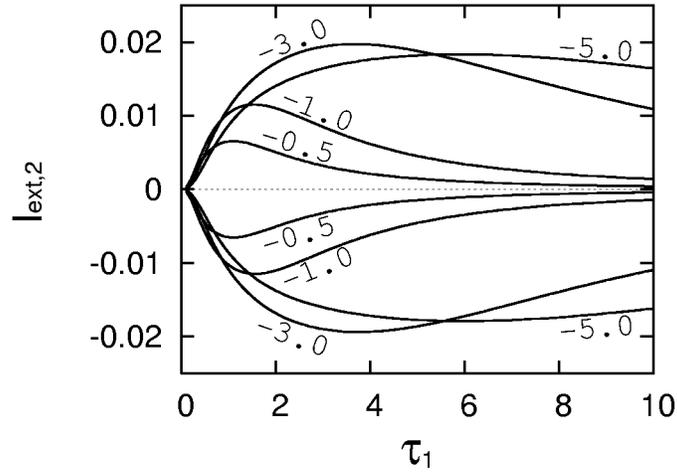}
\end{center}
\caption{The upper and lower bounds of $I_{{{\mbox{\footnotesize {ext}}}},2}$ for
entrainment of two clusters of neurons are plotted against $\tau_1$
($Q=2,1\leq N_1=N_2,\tau_2=0.1\tau_1,$ and $I_{{{\mbox{\footnotesize {ext}}}},1}=0$). The numbers in the
figure indicate the value of $g$.}\label{entrainment}
\end{figure}

We extend the above result to investigate the case when the excitability
of neurons are different between two clusters.  Fixing
$I_{{{\mbox{\footnotesize {ext}}}},1}=0$, we investigate the behavior of $\varphi_2$ with
change of $I_{{{\mbox{\footnotesize {ext}}}},2}$ in {{{Fig.}~\ref{8}}}.  With $I_{{{\mbox{\footnotesize {ext}}}},2}=0$, we
find three stable and two unstable solutions, which are consistent with
the preceding results in {{{Fig.}~\ref{bifurcation}}}. The in-phase solution
$\varphi_2=0,1$ is extended by the change of $I_{{{\mbox{\footnotesize {ext}}}},2}$ within the
interval $-0.019{{\raisebox{-1ex}{\mbox{$\stackrel{{<}}{\sim}$}}}} I_{{{\mbox{\footnotesize {ext}}}},2} {{\raisebox{-1ex}{\mbox{$\stackrel{{<}}{\sim}$}}}} 0.020$.  In this interval
two clusters of neurons show synchronized firing with small phase
difference, that is, entrainment occurs.  To examine the robustness of
this entrainment, we plot this range of $I_{{{\mbox{\footnotesize {ext}}}},2}$ as a function
of $\tau_1$ in {{{Fig.}~\ref{entrainment}}}.  The remarkable feature of this
phase diagram is the narrow range of $I_{{{\mbox{\footnotesize {ext}}}},2}$ with short decay
time constant~$\tau_1$, and it is interesting that the pitchfork bifurcation described in
Fig.~\ref{bifurcation} explains this narrow range of $I_{{{\mbox{\footnotesize {ext}}}},2}$. 
In this bifurcation diagram, the out-of-phase
solutions merge into the in-phase solutions at $\tau_1=0$. Therefore,
the entrained solution in {{{Fig.}~\ref{8}}} vanishes in the limit
$\tau_1\rightarrow 0$, and this vanishment explains the zero range of
$I_{{{\mbox{\footnotesize {ext}}}},2}$ at $\tau_1=0$ in {{{Fig.}~\ref{entrainment}}}.

On the other hand, the out-of-phase solution ($\varphi_2=0.5$) is
considerably robust against the change of $I_{{{\mbox{\footnotesize {ext}}}},2}$, especially
with short $\tau_1$ ($-0.083{\raisebox{-1ex}{\mbox{$\stackrel{{<}}{\sim}$}}}I_{{{\mbox{\footnotesize {ext}}}},2}{\raisebox{-1ex}{\mbox{$\stackrel{{<}}{\sim}$}}}0.080$
with $\tau_1=1.5$.)  Nevertheless, when we apply the external currents
to halves of neurons of both clusters (i.e., $Q=4,r_1=r_2=r_3=r_4=0.25,
I_{{{\mbox{\footnotesize {ext}}}},1}=I_{{{\mbox{\footnotesize {ext,3}}}}}=0,I_{{{\mbox{\footnotesize {ext}}}},2}=I_{{{\mbox{\footnotesize {ext}}}},4}\ne
0,\varphi_1=0,\varphi_2\sim 0,\varphi_3\sim 0.5,\varphi_4\sim 0.5$), the
range for successful entrainment is found to be narrow
($-0.016{\raisebox{-1ex}{\mbox{$\stackrel{{<}}{\sim}$}}}I_{{{\mbox{\footnotesize {ext}}}},2}=I_{{{\mbox{\footnotesize {ext}}}},4}{\raisebox{-1ex}{\mbox{$\stackrel{{<}}{\sim}$}}}0.017$ with
$\tau_1=1.5$).  It seems that cluster synchronization easily breaks when
we apply heterogeneous external electric currents that cause splitting
of clusters.

\section{One-cluster state ($Q=1$) in networks of Hodgkin-Huxley (HH) neurons}\label{sec:hodgkin}

To explore the biological plausibility of synchronization in IF neurons
we study more realistic neuron model that is defined by HH equations.
A HH neuron, whose dynamics is described in
appendix~\ref{sec:appendix-hodgkin}, does not show intrinsic firing without external stimulus.  Therefore, we apply constant external
electric current~$I_{\mbox{ext}}=10~(\mu A/cm^2)$ to all of HH neurons
and analyze synchronization in intrinsically firing homogeneous HH
neurons assuming the same synaptic couplings as Eq.~(\ref{uniform}).
Figure~\ref{fig:hodgkin} shows $\tau_1-g$~phase diagram, where the
condition for stable one-cluster state ($Q=1$ and $2\le N$) is
described.  In the large area of inhibitory couplings ($g<0$) we find
stable in-phase synchronization.  Beyond $\tau_1=7.0$ the behavior of
$|\lambda_{2}^{{\tiny \parallel}}|$ and $|\lambda_{11}^{{\tiny \perp}}|$ is similar to those
of IF neurons described in Fig.~\ref{c1}{(a)}, and the change of stability
occurs at $g=0$ because of $|\lambda_{11}^{{\tiny \perp}}|$.  Below
$\tau_1=7.0$, however, synchronization with inhibitory couplings takes
place only below a certain negative value of $g$, and 
excitatory couplings can induce synchronization in some conditions. $\tau_1-g$~phase
diagram of IF neurons (Fig.~\ref{c1}{(b)}) can explain synchronization
in HH neurons with slowly decaying synapses, though the synchronization
condition of HH neurons with fast decaying synapses is more complicated
than IF neurons.

\section{Discussion}\label{sec:discussion}

\begin{figure}
 
 {}{}{}
\begin{center}
\includegraphics[scale=1.5]{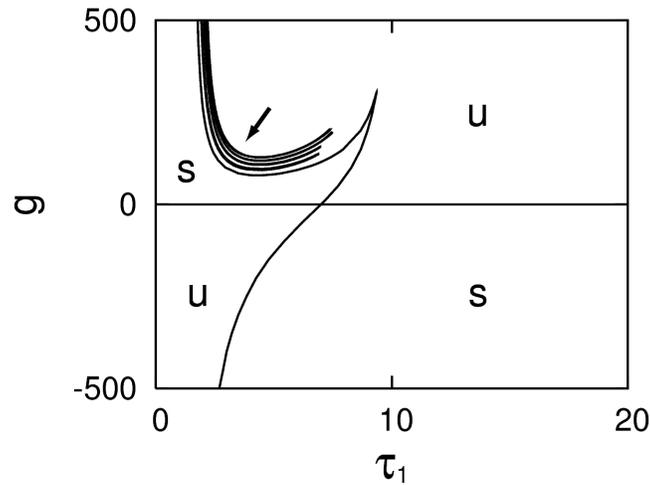}
\end{center}
 \caption{$\tau_1-g$~phase diagram for one-cluster state ($Q=1$) of 
 multiple Hodgkin-Huxley (HH) neurons ($2\le N$) under the condition
 $\tau_2=0.1\tau_1$. ``s'' (``u'') in the figure indicates the region for
 the stable (unstable) one-cluster state. Around the arrow we find a lot of
 isolated regions for the stable one-cluster state. Note that we apply
 constant external electric current~$I_{\mbox{ext}}=10~(\mu A/cm^2)$ to
 all of HH neurons so as to induce intrinsic firing of neurons.}\label{fig:hodgkin}
\end{figure}

We have studied cluster state of networks of spiking neurons.  We have
shown the analytical method that can deal with synchronization in the
large size of neural networks with arbitrary neuron dynamics and
arbitrary interactions.  Employing this analysis we have investigated
networks of IF neurons interconnected through uniform chemical synapses.  In the
analysis of one-cluster state, we have found the
change of stability of a cluster, which has elucidated that in-phase
synchronization of multiple IF neurons occurs only with inhibitory
couplings (Fig.~\ref{c1}).  It must be noted that this analytical result well explains the
structure of interneurons in the real nervous system, where interneurons
are interconnected through inhibitory chemical synapses.  In addition,
we have investigated the entrainment of two clusters of IF neurons with
different excitability (Fig.~\ref{entrainment}). We have explained the
narrow range of $I_{{{\mbox{\footnotesize {ext}}}},2}$ with short decay time constant $\tau_1$
in Fig.~\ref{entrainment} by the bifurcation diagram described in
Fig.~\ref{bifurcation}. Furthermore, we have investigated one-cluster
state of HH neurons.  HH neurons show stable in-phase synchronization in
the large parameter region of inhibitory chemical synapses, though the
synchronization condition of HH neurons with fast decaying synapses is
more complicated than IF neurons (Fig.~\ref{fig:hodgkin}).

\begin{figure}
 {}{}{}
 {}{}{}
 {}{}{}
 {}
\begin{center}
\begin{tabular}{ccc}
{(a)} & & \ \ \ \\
 & \includegraphics[scale=1.3]{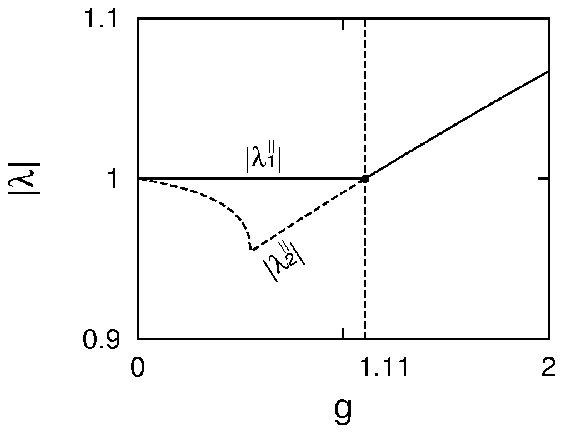} \\
{(b)} \\
 & \includegraphics[scale=1.3]{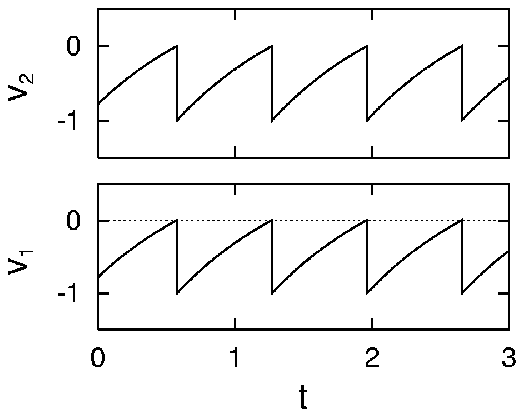} \\
{(c)} \\
 & \includegraphics[scale=1.3]{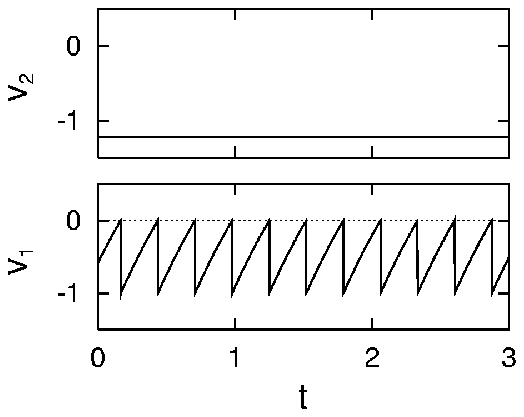} 
\end{tabular}
\end{center}
 \caption{Stability of in-phase synchronization of a couple of IF neurons
 interconnected with $J_{11}=J_{22}=-J_{12}=-J_{21}=g/2$ ($\tau_1=3.5$ and $\tau_2=0.1\tau_1$).
 {(a)}\ Absolute values of $\lambda_1^{{\tiny \parallel}}$ and
 $\lambda_2^{{\tiny \parallel}}$ are plotted as a function of $g$.
 Beyond $g=1.11$, the in-phase synchronization becomes unstable.
 {(b)}\ The result of numerical simulations with $g=1.0$.
 A couple of neurons show in-phase synchronization.
 {(c)}\ The result of numerical simulations with $g=1.2$.
 Only a single neuron fires at high frequency in the winner-take-all fashion.}\label{fig:counterexample}
\end{figure}

Although van~Vreeswijk {\it et al.}  have proposed another type of
stability criterion based on
function~$G(\phi)$\cite{vreeswijk1,vreeswijk2}, this stability criterion
is unsound in some conditions.  One counterexample of their criterion is
a couple of IF neurons with couplings
$J_{11}=J_{22}=-J_{12}=-J_{21}=g/2$.  With $\tau_1=3.5$ and
$\tau_2=0.1\tau_1$, in-phase synchronization of these neurons becomes
unstable beyond the critical point~$g=1.11$ as shown in
Fig.~\ref{fig:counterexample}.  While our analysis based on linear
stability precisely yields this critical point, van~Vreeswijk's
criterion, namely, $G(\phi)=-g{\left ( {e^{-T}/2} \right )}\int_0^1 e^{T\theta}
{\left ( {\tilde{S}{\left [ {T{\left ( {\theta+\phi} \right )}} \right ]}-\tilde{S}{\left [ {T{\left ( {\theta-\phi} \right )}} \right ]}} \right )}d\theta$
with $T=\log{\left ( {v_0-v_r/v_0} \right )}$, fails to give the critical point.
Gerstner {\it et al.} have also investigated networks of IF
neurons\cite{gerstner}.
Their analysis, however, cannot treat the
realistic form of synaptic electric current~$S(t)$ that exerts the
long-time influence after activation within the finite size of matrix.

The present decomposition of linear stability is simple enough to
investigate the general neuron dynamics including FN neurons and HH
neurons. Even when the behavior of neurons are chaotic\cite{feudel}, we
are still able to evaluate the stability of cluster state using tangential
Lyapunov exponents and transversal Lyapunov exponents\cite{myosioka8}, and such technique
may give a deeper understanding of the complicated behavior of
HH neurons around the arrow in Fig.~\ref{fig:hodgkin}.  It is
interesting to apply the present analysis to networks including
pyramidal neurons as well as interneurons\cite{kopell}.  The surface of
the neocortex is subdivided into numerous columnar organizations, each
of which is composed of several layers of neurons \cite{mountcastle}.
The internal and external dynamics of such columnar organizations would
also be the future target of the present analysis.

\appendix

\section{The Hodgkin-Huxley (HH) equations}\label{sec:appendix-hodgkin}

The HH equations are the four-dimensional ordinary differential
equations, which describe the spike generation of the squid's giant
axon\cite{hodgkin}.  The dynamics of a neuron state
vector~${{\bf x}}={\left ( {v,w_{1},w_{2},w_{3}} \right )}^{{\tiny\mbox{T}}}$ for a HH neuron is expressed as
\begin{eqnarray}
C_m\ \dot{v}&=&\overline{g}_{Na}w_2^3w_1{\left ( {v_{Na}-v} \right )}+\overline{g}_{K}w_3^4{\left ( {v_{K}-v} \right )}\nonumber\\
&&+\overline{g}_{L}{\left ( {v_{L}-v} \right )}+I_{\mbox{ext}},\\
\dot{w}_1&=&\alpha_1{\left ( {1-w_1} \right )}-\beta_1 w_1,\\
\dot{w}_2&=&\alpha_2{\left ( {1-w_2} \right )}-\beta_2 w_2,\\
\dot{w}_3&=&\alpha_3{\left ( {1-w_3} \right )}-\beta_3 w_3
\end{eqnarray}
with
\begin{eqnarray}
\alpha_1&=&0.01{\left ( {10-v} \right )}\left /{\left\{ {\exp{\left ( {\frac{10-v}{10}} \right )}-1} \right \}}\right . ,\\
\beta_1&=&0.125\exp{\left ( {-v/80} \right )},\\
\alpha_2&=&0.1{\left ( {25-v} \right )}\left /{\left\{ {\exp{\left ( {\frac{25-v}{10}} \right )}-1} \right \}}\right . ,\\
\beta_2&=&4\exp{\left ( {-v/18} \right )},\\
\alpha_3&=&0.07\exp{\left ( {-v/20} \right )},\\
\beta_3&=&1\left /{\left\{ {\exp{\left ( {\frac{30-v}{10}} \right )}-1} \right \}}\right .,
\end{eqnarray}
where $v_{Na}=50\mbox{ [mV]},\
v_K=-77\mbox{ [mV]},\ v_L=-54.4\mbox{ [mV]},\
\overline{g}_{Na}=120\mbox{ [mS/cm$^2$]},\ \overline{g}_K=36\mbox{
[mS/cm$^2$]},\ \overline{g}_L=0.3\mbox{ [mS/cm$^2$]},\ {\mbox{ and }} C_m=1\mbox{
[$\mu$F/cm$^2$]}.$
In the present study we set $I_{\mbox{ext}}=10~(\mu A/cm^2)$ to induce intrinsic firing of a HH neuron.

\begin{thebibliography}{99}
\bibitem{buzsaki3} G.~Buzs{\'a}ki, Z.~Horv{\'a}th, R.~Urioste, J.~Hetke,
	and K.~Wise, Science, 256, 1025 (1992).
\bibitem{wang} X.J.~Wang and G.~Buz{\'a}ki, J. Neurosci., 16, 6402 (1996).
\bibitem{ermentrout} G.B.~Ermentrout and N.~Kopell,
	SIAM J. Math. Anal., 15, 215 (1984).
\bibitem{hansel} D.~Hansel, G.~Mato, and C.~Meunier, Neural Comp., 7,
	307 (1995).
\bibitem{bressloff3} P.C.~Bressloff and S.~Coombes, Neural Comp., 12, 91 (2000).
\bibitem{kuramoto} Y.~Kuramoto, Chemical oscillations, waves, and
		   turbulence (Springer-Verlag 1984).
\bibitem{fujisaka} H.~Fujisaka and T.~Yamada, Prog. Theor. Phys. 69, 32 (1983).
\bibitem{kaneko} K.~Kaneko, Physica D, 77, 456 (1994).
\bibitem{maistrenko} Y.~Maistrenko and T~Kapitaniak, Phys. Rev. E, 54, 3285
        (1996).
\bibitem{pikovsky} A.~Pikovsky, O.~Popovych, and Yu.~Maistrenko, Phys. Rev. Lett., 87, 044102
	(2001).
\bibitem{myosioka5} M.~Yoshioka, Phys. Rev. E, 65, 011903 (2002).
\bibitem{myosioka6} M.~Yoshioka, Phys. Rev. E, 66, 061913 (2002).
\bibitem{hodgkin} A.L.~Hodgkin and A.F.~Huxley, J. Physiol., 117, 500 (1952).
\bibitem{vreeswijk1} C.~van Vreeswijk, L.F.~Abbott, and G.B.~Ermentrout,
	J. Comp. Neurosci, 1, 313 (1994).
\bibitem{vreeswijk2} C.~van~Vreeswijk, Phys. Rev. E 54, 5522 (1996).
\bibitem{gerstner} W.~Gerstner, J.L.~van~Hemmen, and J.D.~Cowan, Neural Comp., 8, 1653 (1996).
\bibitem{feudel} U.~Feudel, A.~Neiman, X.~Pei, W.~Wojtenek, H.~Braun,
	M.~Huber, and F.~Moss, Chaos, 10, 231 (2000).
\bibitem{myosioka8} M.~Yoshioka, Phys. Rev. E, in press.
\bibitem{kopell} N.~Kopell, G.B.~Ermentrout, M.A.~Whittington, and
	R.D.~Traub, Proc. Natl. Acad. Sci. USA, 97, 1867 (2000)
\bibitem{mountcastle} V.B.~Mountcastle, J. Neurophysiol. 20, 408 (1957)
\end{thebibliography}
\end{document}